\documentclass[aps,prl,twocolumn,showpacs,groupedaddress]{revtex4}

\usepackage{graphicx}

\begin{document}

\newcommand{\beq}{\begin{equation}}
\newcommand{\eeq}{\end{equation}}
\newcommand{\barr}{\begin{eqnarray}}
\newcommand{\earr}{\end{eqnarray}}

\newcommand{\andy}[1]{ }
\def\cH{{\cal H}}
\def\cV{{\cal V}}
\def\cU{{\cal U}}
\def\bra#1{\langle #1 |}
\def\ket#1{| #1 \rangle}


\title{Quantum Zeno subspaces}


\author{P. Facchi}
\author{S. Pascazio}
\affiliation{Dipartimento di Fisica, Universit\`a di Bari  I-70126 Bari,
Italy}
\affiliation{Istituto Nazionale di Fisica Nucleare, Sezione di Bari,
I-70126 Bari, Italy }


\date{\today}

\begin{abstract}
The quantum Zeno effect is recast in terms of an adiabatic theorem
when the measurement is described as the dynamical coupling to
another quantum system that plays the role of apparatus. A few
significant examples are proposed and their practical relevance
discussed. We also focus on decoherence-free subspaces.
\end{abstract}


\maketitle


If very frequent measurements are performed on a quantum system,
in order to ascertain whether it is still in its initial state,
transitions to other states are hindered and the quantum Zeno
effect takes place \cite{Beskow,Misra}. This phenomenon stems from
general features of the Schr\"odinger equation that yield
quadratic behavior of the survival probability at short times
\cite{strev,PIO}.
The first realistic test of the quantum Zeno effect (QZE) for
oscillating (two-level) systems was proposed about 15 years ago
\cite{Cook}. This led to experiments, discussions and new
proposals \cite{Itano}.
A few years ago, the presence of a short-time quadratic region was
experimentally confirmed also for a \textit{bona fide} unstable
system \cite{Wilkinson}. The same experimental setup has been used
very recently \cite{raizenlatest} in order to prove the existence
of the Zeno effect (as well as its inverse
\cite{antiZeno,heraclitus}) for an unstable quantum mechanical
system, leading to new ideas \cite{Agarwal01,Frishman01}.

It is important to stress that the quantum Zeno effect does not
necessarily freeze everything. On the contrary, for frequent
projections onto a multi-dimensional subspace, the system can
evolve away from its initial state, although it remains in the
subspace defined by the measurement. This continuing time
evolution within the projected subspace (``quantum Zeno dynamics")
has been recently investigated
\cite{compactregularize}. It has peculiar physical and mathematical
features and sheds light on some subtle mathematical issues
\cite{Friedman72,Gustafson,Misra}.

All the above-mentioned investigations deal with what can be
called ``pulsed" measurements, according to von Neumann's
projection postulate \cite{von}. However, from a physical point of
view, a ``measurement" is nothing but an interaction with an
external system (another quantum object, or a field, or simply
another degree of freedom of the very system investigated),
playing the role of apparatus. In this respect, if one is not too
demanding in philosophical terms, von Neumann's postulate can be
regarded as a useful shorthand notation, summarizing the final
effect of the quantum measurement. This simple observation enables
one to reformulate the QZE in terms of a (strong) coupling to an
external agent. We emphasize that in such a case the QZE is a
consequence of the dynamical features (i.e.\ the form factors) of
the coupling between the system investigated and the external
system, and no use is made of projection operators (and
non-unitary dynamics). The idea of ``continuous" measurement in a
QZE context has been proposed several times during the last two
decades
\cite{Peres80,PeresKraus}, although the first quantitative
comparison with the ``pulsed" situation is rather recent
\cite{Schulman98}.

The purpose of the present article is to cast the quantum Zeno
evolution in terms of an adiabatic theorem and study possible
applications. We will see that the evolution of a quantum system
is profoundly modified (and can be tailored in an interesting way)
by a continuous measurement process: the system is forced to
evolve in a set of orthogonal subspaces of the total Hilbert space
and a dynamical superselection rule arises in the strong coupling
limit. These general ideas will be corroborated by some simple
examples.

\begin{figure}
\includegraphics[height=7.4cm]{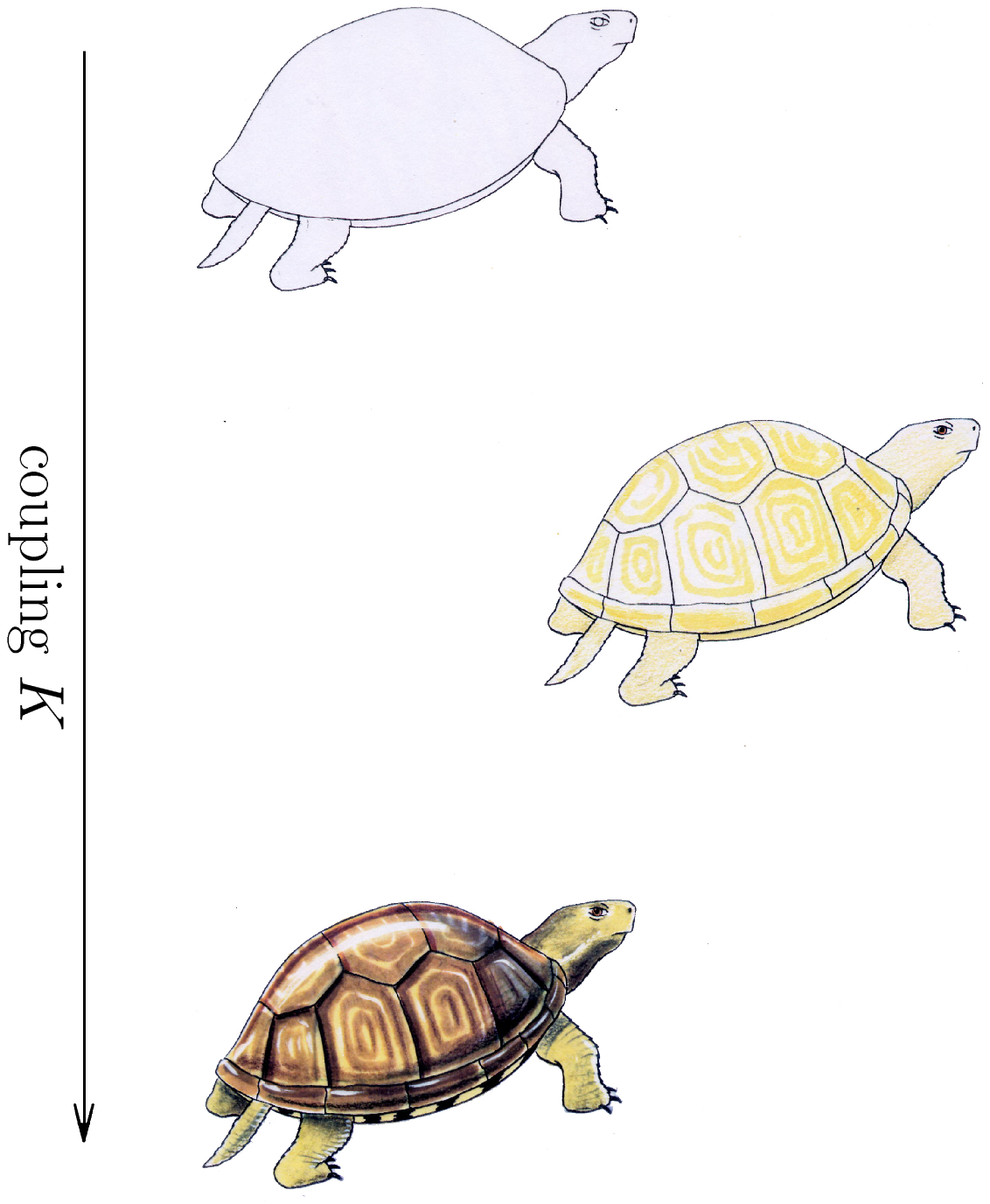}
\caption{\label{tortoise} The Hilbert space of the system:
an effective superselection rule appears as the coupling $K$ to
the apparatus is increased.}
\end{figure}

We start by considering the case of ``pulsed" observation. We
first extend Misra and Sudarshan's theorem
\cite{Misra} in order to accomodate multiple projectors. Let Q be
a quantum system, whose states belong to the Hilbert space ${\cal
H}$ and whose evolution is described by the unitary operator
$U(t)=\exp(-iHt)$, where $H$ is a time-independent lower-bounded
Hamiltonian. Let $\{P_n\}_n \, (P_nP_m=\delta_{mn}P_n, \sum_n
P_n=1)$ be a (countable) collection of projection operators and
$\text{Ran}P_n={\cal H}_{P_n}$ the relative subspaces. This
induces a partition on the total Hilbert space $\cH=\bigoplus_n
\cH_{P_n}$. Let $\rho_0$ be the initial density matrix of the
system. We ``prepare" the system by performing an initial
measurement, described by the superoperator
\beq
\label{eq:superP}
\hat P \rho=\sum_n P_n \rho P_n = \rho_0.
\eeq
The free evolution reads
\beq
\hat U_t \rho_0=U(t) \rho_0 U^\dagger(t),\qquad
U(t)=\exp(-i H t)
\eeq
and the Zeno evolution after $N$ measurements in a time $t$ is
governed by the superoperator
\beq
\hat V^{(N)}_t=\hat P\left(\hat U \left(t/N \right)\hat
P\right)^{N-1} ,
\eeq
which yields
\beq
\rho(t)=\hat V^{(N)}_t \rho_0
=\sum_{n_1,\dots,n_N}V_{n_1\dots n_N}^{(N)}(t)\; \rho_0\;
V_{n_1\dots n_N}^{(N)\dagger}(t) ,
\eeq
where
\barr
V_{n_1\dots n_N}^{(N)}(t)  = P_{n_N} U\left(t/N\right) P_{n_{N-1}}
\cdots P_{n_2} U\left(t/N\right) P_{n_1}.
\earr
We follow \cite{Misra} and assume for each $n$ the existence of
the strong limits ($t>0$)
\andy{slims}
\beq
\lim_{N\to\infty} V_{n\dots n}^{(N)}(t) \equiv \cV_n (t), \qquad
\lim_{t \rightarrow 0^+} {\cal V}_n(t) = P_n .
\label{eq:slims}
\eeq
Then ${\cal V}_n(t)$ exist for all real $t$ and form a semigroup
\cite{Misra}, and $\cV_n^\dagger(t)\cV_n(t)=P_n$. Moreover it is easy
to show that
\beq
\lim_{N\to\infty} V_{n\dots n'\dots}^{(N)}(t) = 0, \qquad
\text{for}\quad n'\neq n .
\eeq
Therefore the final state is
\andy{rhoZ}
\barr
& & \rho(t)=\hat \cV_t\rho_0
=\sum_n \cV_n(t) \rho_0 \cV_n^\dagger(t), \label{eq:rhoZ} \\
& & \qquad
\text{with} \quad \sum_n \cV_n^\dagger(t)\cV_n(t)=\sum_n P_n=1 .
\nonumber \earr
The components $\cV_n(t) \rho_0 \cV_n^\dagger(t)$ make up a block
diagonal matrix: the initial density matrix is reduced to a
mixture and any interference between different subspaces
$\cH_{P_n}$ is destroyed (complete decoherence). In conclusion,
\andy{probinfu}
\beq
p_n(t) =  \text{Tr} \left(\rho(t) P_n\right)=
\text{Tr}\left(\rho_0 P_n\right)=p_n(0) , \quad \forall n .
\label{eq:probinfu}
\eeq
In words, probability is conserved in each subspace and no
probability ``leakage" between any two subspaces is possible. The
total Hilbert space splits into invariant subspaces and the
different components of the wave function (or density matrix)
evolve independently within each sector. One can think of the
total Hilbert space as the shell of a tortoise, each invariant
subspace being one of the scales. Motion among different scales is
impossible. (See Fig.\ \ref{tortoise} in the following.) The study
of the Zeno dynamics \textit{within} a given infinite-dimensional
subspace is an interesting problem
\cite{compactregularize} that will not be discussed here. The
original formulation of the Zeno effect is reobtained when $p_n=1$
for some $n$, in (\ref{eq:probinfu}): the initial state is then in
one of the invariant subspaces and the survival probability in
that subspace remains unity.

The previous theorem hinges upon von Neumann's projections
\cite{von}. However, as we explained in the introduction, a QZE
can also be obtained by performing a continuous measurement on a
system. For example, consider the Hamiltonian
\andy{ham3l}
\beq
H_{\text{3lev}} = \Omega \sigma_1+ K \tau_1=
\pmatrix{0 &
\Omega & 0\cr \Omega & 0 & K \cr 0 & K & 0},
\label{ham3l}
\eeq
describing two levels (system), with Hamiltonian $H=\Omega
\sigma_1=\Omega (\ket{1}\bra{2}+ \ket{2}\bra{1})$, coupled to a
third one, that plays the role of measuring apparatus: $K
H_{\rm\text{ meas}}= K \tau_1=K (\ket{2}\bra{3}+ \ket{3}\bra{2})$.
This model, first considered in \cite{Peres80}, is probably the
simplest way to include an ``external" apparatus in our
description: as soon as the system is in $|2\rangle$ it undergoes
Rabi oscillations to $|3\rangle$. We expect level $|3\rangle$  to
perform better as a measuring apparatus when the strength $K$ of
the coupling becomes larger. Indeed, if initially the system is in
state $\ket{1}$, the survival probability reads
\andy{sp3}
\beq
p(t)= \left[K^2+ \Omega^2 \cos(K_1 t) \right]^2/K_1^4\;
\stackrel{K\to \infty}{\longrightarrow} 1,
\label{sp3}
\eeq
where $K_1=\sqrt{K^2+\Omega^2}$. This simple model captures many
interesting features of a Zeno dynamics (and will help clarify our
general approach). Many similar examples can be considered: in
general
\cite{PIO,Napoli}, one can include the detector in the quantum
description, by considering the Hamiltonian
\beq
\label{eq:sys+meas}
H_K=H+ K H_{\text{meas}},
\eeq
where $H$ is the Hamiltonian of the system under observation (and
can include the free Hamiltonian of the apparatus) and
$H_{\text{meas}}$ is the interaction Hamiltonian between the
system and the apparatus.

We now prove a theorem, which is the exact analog of Misra and
Sudarshan's theorem for a dynamical evolution of the type
(\ref{eq:sys+meas}). Consider the time evolution operator
\barr
U_{K}(t) = \exp(-iH_K t) .
\label{eq:measinter}
\earr
We will prove that in the ``infinitely strong measurement" limit
$K\to\infty$ the evolution operator
\beq
\label{eq:limevol}
\cU(t)=\lim_{K\to\infty}U_{K}(t),
\eeq
becomes diagonal with respect to $H_{\text{meas}}$:
\beq
\label{eq:diagevol}
[\cU(t), P_n]=0, \quad\text{where}\quad H_{\text{meas}} P_n=\eta_n
P_n,
\eeq
$P_n$ being the orthogonal projection onto $\cH_{P_n}$, the
eigenspace of $H_{\text{meas}}$ belonging to the eigenvalue
$\eta_n$. Note that in Eq.\ (\ref{eq:diagevol}) one has to
consider distinct eigenvalues, i.e., $\eta_n\neq\eta_m$ for $n\neq
m$, whence the $\cH_{P_n}$'s are in general multidimensional.

The theorem is easily proven by recasting it in the form of an
adiabatic theorem. In the $H$ interaction picture,
\beq
H^{\text{I}}_{\text{meas}}(t)=e^{i H t} H_{\text{meas}} e^{-i H t}
,
\eeq
the Schr\"odinger equation reads
\beq
i \partial_t U_K^{\text{I}} (t) = K H^{\text{I}}_{\text{meas}}(t)
U_K^{\text{I}} (t) .
\eeq
This has exactly the same form of an adiabatic evolution $i
\partial_s U_T (s) = T H(s) U_T (s)$ \cite{Messiah61}: the
large coupling $K$ limit corresponds to the large time $T$ limit
and the physical time $t$ to the scaled time $s=t/T$. In the
$K\to\infty$ limit, by considering a spectral projection
$P^{\text{I}}_n(t)=e^{i H t} P_n e^{-i H t}$ of
$H^{\text{I}}_{\text{meas}}(t)$, the limiting operator ${\cal
U}^{\text{I}} (t)=\lim_{K\to\infty}U_K^{\text{I}} (t)$ satisfies
the intertwining property ${\cal U}^{\text{I}} (t)
P^{\text{I}}_n(0)=P^{\text{I}}_n(t){\cal U}^{\text{I}} (t)$, i.e.\
maps $\cH_{P^{\text{I}}_n(0)}$ onto $\cH_{P^{\text{I}}_n(t)}$:
\beq
\psi^{\text{I}}_0\in \cH_{P^{\text{I}}_n(0)}\rightarrow
\psi^{\text{I}}(t)\in \cH_{P^{\text{I}}_n(t)}.
\eeq
In the Schr\"odinger picture
\beq
\psi_0\in \cH_{P_n}\rightarrow \psi(t)\in \cH_{P_n},
\eeq
whence
\barr
\rho(t)= e^{-i H t}\cU^{\text{I}}(t) \rho_0
\cU^{{\text{I}}\dagger}(t)e^{i H t}=\cU(t) \rho_0 \cU^\dagger(t) ,
\earr
where $\cU(t)$ has the property (\ref{eq:diagevol}) and the
probability to find the system in $\cH_{P_n}$ satisfies Eq.\
(\ref{eq:probinfu}) and is therefore constant: if the initial
state of the system belongs to a given sector, it will be forced
to remain there forever (QZE).

Even more, by exploiting the features of the adiabatic theorem in
greater details, it is possible to show that, for time independent
Hamiltonians, the limiting evolution operator has the explicit
form \cite{tokyo}
\beq
\label{eq:theorem}
\cU(t)=\exp[-i(H_{\text{diag}}+K H_{\text{meas}}) t],
\eeq
where
\beq
H_{\text{diag}}=\sum_n P_n H P_n
\label{eq:diagsys}
\eeq
is the diagonal part of the system Hamiltonian $H$ with respect to
the interaction Hamiltonian $H_{\text{meas}}$.

Let us briefly comment on the physical meaning. In the
$K\to\infty$ limit, due to (\ref{eq:diagevol}), the time evolution
operator becomes diagonal with respect to $H_{\text{meas}}$,
$[\cU(t), H_{\text{meas}}]=0$, an effective superselection rule
arises and the total Hilbert space is split into subspaces
$\cH_{P_n}$ that are invariant under the evolution. These
subspaces are defined by the $P_n$'s, i.e., they are eigenspaces
belonging to distinct eigenvalues $\eta_n$: in other words,
subspaces that the apparatus is able to distinguish. On the other
hand, due to (\ref{eq:diagsys}), the dynamics within each Zeno
subspace $\cH_{P_n}$ is governed by the diagonal part $P_n H P_n$
of the system Hamiltonian $H$. This bridges the gap with the
description (\ref{eq:superP})-(\ref{eq:probinfu}) and clarifies
the role of the detection apparatus. In Fig.\ \ref{tortoise} we
endeavored to give a pictorial representation of the decomposition
of the Hilbert space as $K$ is increased. It is worth noticing
that the superselection rules discussed here are \textit{de facto}
equivalent to the celebrated ``W$^3$" ones \cite{WWW}, but turn
out to be a mere consequence of the Zeno dynamics.

Four examples will prove useful. First example: reconsider
$H_{\text{3lev}}$ in Eq.\ (\ref{ham3l}). As $K$ is increased, the
Hilbert space is split into three invariant subspaces (the three
eigenspaces of $H_{\text{meas}}=\tau_1$): (level $\ket{1}$)
$\oplus$ (level $\ket{2}+\ket{3}$) $\oplus$ (level
$\ket{2}-\ket{3}$).

Second example: consider
\andy{ham4l}
\beq
H_{\text{4lev}} =
\pmatrix{0 &
\Omega & 0 & 0 \cr \Omega & 0 & K & 0 \cr 0 & K & 0 & K'
\cr 0 & 0 & K' & 0 },
\label{ham4l}
\eeq
where level $\ket{4}$ ``measures" whether level $\ket{3}$ is
populated. If $K' \gg K \gg \Omega$, the total Hilbert space is
divided into three subspaces: (levels $\ket{1}$ and $\ket{2}$)
$\oplus$ (level $\ket{3}+\ket{4}$) $\oplus$ (level
$\ket{3}-\ket{4}$). Notice that the $\Omega$ oscillations are
\textit{restored} as $K' \gg K$ (in spite of $K \gg \Omega$). A
watched cook can freely watch a boiling pot.

Third example (decoherent-free subspaces \cite{Viola99} in quantum
computation). The Hamiltonian \cite{Beige00}
\beq
\label{eq:cavity}
H_{\text{meas}}=i g \sum_{i=1}^2 [ b\; \ket{2}_{ii}\bra{1} -
\text{H.c.}] - i \kappa b^\dagger b
\eeq
describes a system of two ($i=1,2$) three-level atoms in a cavity.
The atoms are in a $\Lambda$ configuration with split ground
states $\ket{0}_i$ and $\ket{1}_i$ and excited state $\ket{2}_i$,
while the cavity has a single resonator mode $b$ in resonance with
the atomic transition 1-2. Spontaneous emission inside the cavity
is neglected, but a photon leaks out through the nonideal mirrors
with a rate $\kappa$. The (5-dimensional) eigenspace $\cH_{P_0}$
of $H_{\text{meas}}$ belonging to the eigenvalue $\eta=0$ is
spanned by
\andy{decfree}
\beq
\{\ket{000},\ket{001},\ket{010},\ket{011},(\ket{021}-\ket{012})/\sqrt{2}\},
\label{decfree}
\eeq
where $\ket{0j_1 j_2}$ denotes a state with no photons in the
cavity and the atoms in state $\ket{j_1}_1\ket{j_2}_2$. If the
coupling $g$ and the cavity loss $\kappa$ are sufficiently strong,
any other weak Hamiltonian $H$ added to (\ref{eq:cavity}) reduces
to $P_0 H P_0$ and changes the state of the system only within the
decoherence-free subspace (\ref{decfree}).

Fourth example. Let
\andy{hamqc}
\beq
H=
\pmatrix{0 &
\tau_{\text{Z}}^{-1} & 0\cr \tau_{\text{Z}}^{-1} & -i
2/ \tau_{\text{Z}}^2 \gamma & K \cr 0 & K & 0}.
\label{hamqc}
\eeq
This describes the spontaneous emission $\ket{1}\to\ket{2}$ of a
system into a (structured) continuum, while level $\ket{2}$ is
resonantly coupled to a third level $\ket{3}$ \cite{PIO}. This
case is also relevant for quantum computation, if one is
interested in protecting a given subspace (level $\ket{1}$) from
decoherence
\cite{Viola99,Beige00} by inhibiting spontaneous emission \cite{Agarwal01}.
Here $\gamma$ represents the decay rate to the continuum and
$\tau_{\text{Z}}$ is the Zeno time (convexity of the initial
quadratic region). As the Rabi frequency $K$ is increased one is
able to hinder spontaneous emission from level $\ket{1}$ (to be
protected) to level $\ket{2}$. However, in order to get an
effective ``protection" of level $\ket{1}$, one needs $K >
1/\tau_{\text{Z}}$. More to this, when the presence of the inverse
Zeno effect is taken into account, this requirement becomes even
more stringent \cite{heraclitus} and yields $K >
1/\tau_{\text{Z}}^2\gamma$. Both these conditions can be very
demanding for a real system subject to dissipation
\cite{PIO,Napoli,heraclitus}. For instance,
typical values for spontaneous decay in vacuum are $\gamma\simeq
10^9$s$^{-1}$, $\tau_{\text{Z}}^2\simeq 10^{-29}$s$^2$ and
$1/\tau_{\text{Z}}^2\gamma\simeq 10^{20}$s$^{-1}$
\cite{hydrogen}.

The formulation of a Zeno dynamics in terms of an adiabatic
theorem is powerful. Indeed one can use all the machinery of
adiabatic theorems in order to get results in this context. An
interesting extension would be to consider time-dependent
measurements
\beq
H_{\text{meas}}=H_{\text{meas}}(t),
\eeq
whose spectral projections $P_n=P_n(t)$ have a nontrivial time
evolution. In this case, instead of confining the quantum state to
a fixed sector, one can transport it along a given path (subspace)
$\cH_{P_n(t)}$. One then obtains a dynamical generalization of the
process pioneered by Von Neumann in terms of projection operators
\cite{von,AV}.
The influence of non-adiabatic corrections, for $K$ large but
finite, as well as practical estimates and protection from
decoherence effects will be considered in a future article.

\begin{acknowledgments}
We thank L.\ Neglia and N.\ Cillo for the drawing.
\end{acknowledgments}


\end{document}